\documentclass[aps,prb,a4paper,twocolumn,10pt,superscriptaddress,notitlepage]{revtex4-2}

\newcommand{\papertitle}{Terahertz nanoscopy of quantum-geometric photovoltaics in bilayer graphene}

\usepackage[utf8]{inputenc}
\usepackage{graphicx}
\usepackage{wrapfig}
\usepackage{amsmath}
\usepackage{amssymb}
\usepackage{caption}
\usepackage{natbib}
\usepackage{float}
\usepackage{tikz}
\usepackage{stfloats}
\usepackage{siunitx}
\usepackage{dcolumn}
\usepackage{lmodern}
\usepackage[T1]{fontenc}
\usepackage[unicode=true,
	    colorlinks=true,
	    linkcolor=blue,
	    citecolor=blue,
	    urlcolor=blue]{hyperref} 
\usepackage{sansmath}
\usepackage{caption}
\usepackage{blindtext}
\usepackage[normalem]{ulem}

\usepackage{chemformula}
\usepackage{empheq}
\usepackage{comment}
\usepackage{bm}
\usepackage{gensymb}

\usepackage{caption}
\usepackage[misc]{ifsym}
\usepackage{fontawesome}

\usepackage{lineno}

\setcitestyle{super}

\usepackage{xr}
\makeatletter

\newcommand*{\addFileDependency}[1]{
\typeout{(#1)}
\@addtofilelist{#1}
\IfFileExists{#1}{}{\typeout{No file #1.}}
}\makeatother

\newcommand*{\myexternaldocument}[1]{%
\externaldocument{#1}%
\addFileDependency{#1.tex}%
\addFileDependency{#1.aux}%
}

\myexternaldocument{supplement}

\usepackage[top=2.5cm,bottom=2.5cm,left=1.5cm,right=1.5cm]{geometry}

\newcommand{\equalcontrib}{\textsuperscript{\normalfont\#}}

\begin{document}

\title{\Large\textsf{\papertitle}}

\author{Bogdan Borodin\equalcontrib}
\email{bborodin@nd.edu}
\affiliation{\footnotesize Department of Physics and Astronomy, University of Notre Dame, Notre Dame, IN 46556,~USA}
\affiliation{\footnotesize Stavropoulos Center for Complex Quantum Matter, University of Notre Dame, Notre Dame, IN 46556,~USA}

\author{Jiangxu Li\equalcontrib}
\affiliation{\footnotesize Department of Physics, University of Texas at Dallas, Richardson, TX 75080,~USA}

\author{~Xuanpu Zhang}
\affiliation{\footnotesize Department of Physics and Astronomy, University of Notre Dame, Notre Dame, IN 46556,~USA}
\affiliation{\footnotesize Stavropoulos Center for Complex Quantum Matter, University of Notre Dame, Notre Dame, IN 46556,~USA}

\author{Aleksei Karaborchev}
\affiliation{\footnotesize Department of Physics and Astronomy, University of Notre Dame, Notre Dame, IN 46556,~USA}
\affiliation{\footnotesize Stavropoulos Center for Complex Quantum Matter, University of Notre Dame, Notre Dame, IN 46556,~USA}

\author{Som Subhra Guria}
\affiliation{\footnotesize Department of Physics and Astronomy, University of Notre Dame, Notre Dame, IN 46556,~USA}
\affiliation{\footnotesize Stavropoulos Center for Complex Quantum Matter, University of Notre Dame, Notre Dame, IN 46556,~USA}

\author{Kenji Watanabe}
\affiliation{\footnotesize Research Center for Electronic and Optical Materials, National Institute for Materials Science, 1-1 Namiki, Tsukuba 305-0044,~Japan}

\author{Takashi Taniguchi}
\affiliation{\footnotesize Research Center for Materials Nanoarchitectonics, National Institute for Materials Science,  1-1 Namiki, Tsukuba 305-0044,~Japan}

\author{Roshan Krishna Kumar}
\affiliation{\footnotesize Catalan Institute of Nanoscience and Nanotechnology (ICN2), CSIC and BIST, Campus UAB, Bellaterra, Barcelona 08193, Spain}

\author{Fan Zhang}
\email{zhang@utdallas.edu}
\affiliation{\footnotesize Department of Physics, University of Texas at Dallas, Richardson, TX 75080,~USA}

\author{Petr Stepanov}
\email{pstepano@nd.edu}
\affiliation{\footnotesize Department of Physics and Astronomy, University of Notre Dame, Notre Dame, IN 46556,~USA}
\affiliation{\footnotesize Stavropoulos Center for Complex Quantum Matter, University of Notre Dame, Notre Dame, IN 46556,~USA}



\begin{abstract}

The bulk photovoltaic effect (BPVE) provides a direct link between quantum geometry and nonlinear light-matter interactions, yet conventional far-field measurements cannot disentangle BPVE from contact- and interface-induced photothermoelectric signals. Here we develop cryogenic, polarization-sensitive terahertz microscopy with deep-subwavelength resolution and apply it to dual-gated AB bilayer graphene. Near resonance between the band gap and terahertz photon energy, we observe enhanced optical conductivity and spatially resolved BPVE. Polarization-dependent measurements reveal a characteristic twofold angular symmetry and a distinct shift-vector direction, both governed by crystalline symmetries and quantum geometry. The evolution of BPVE with carrier density and temperature provides a sensitive probe of the fine electronic structure of bilayer graphene, revealing signatures of trigonal warping and electron-hole asymmetry. Agreement with theory demonstrates that the terahertz nanoscopy of BPVE established here offers an optical lens into the interplay between band and quantum geometry in quantum materials.

\begin{quotation}
    \noindent\hspace*{-2.5em}%
    \small \textsuperscript{\#}These authors contributed equally to this work.
\end{quotation}

\end{abstract}


\maketitle  

\vspace{-2ex}
Photocurrent generation in homogeneous crystals without external bias or interfaces was discovered in the 1970s, a phenomenon now known as the bulk photovoltaic effect (BPVE)\cite{10.1063/1.1655453}. Subsequent theoretical developments established a shift-vector picture and a symmetry-based description of polarization-dependent photocurrents\cite{belinicher1980photogalvanic, ivchenko1978new, belinicher1982kinetic, ivchenko1984linear, sipe2000second}. 
Later, the development of Berry-phase theory of modern electronic polarization provided a unified geometric framework for understanding these phenomena\cite{king1993theory, resta1994macroscopic,young2012first}. 
Today, the BPVE and related nonlinear optical responses are recognized as manifestations of both band and quantum geometries, providing powerful probes of the electronic structures of quantum materials including both energy bands and wave functions\cite{ma2021topology, ma2022intelligent, ahn2022riemannian, krishna2025terahertz, gao2025quantum}.

A central quantity in this description is the quantum geometric tensor~\cite{gao2025quantum,ma2021topology},
\begin{equation}
Q_{ab}=g_{ab}+\frac{i}{2}\Omega_{ab},
\end{equation}
where $g_{ab}$ and $\Omega_{ab}$ are the quantum metric and Berry curvature, respectively. While Berry curvature has long been experimentally accessible through transport measurements, direct probes of the quantum metric remain comparatively challenging\cite{rhim2020quantum, tian2023evidence}. 
Intriguingly, the BPVE provides a promising route: 
for a two-band system under an optical electric field polarized along the $b$ direction, the shift current along the $a$ direction at photon energy $\hbar\omega$ is determined by
\begin{equation}
\sigma_{abb}(\omega)
=
\frac{\pi e^3}{\hbar^2}
D(\omega)
\left\langle
g_{bb}(\bm{k})R_a(\bm{k})
\right\rangle_{\omega}.\label{sigma} 
\end{equation}
Here, the joint density-of-state (JDOS) $D(\omega)$ measures the amount of $\bm k$-space available for the optical transition, the quantum metric $g_{bb}$ is the optical transition strength, the shift vector $\bm R$, i.e., the gauge-invariant vector field of the Berry curvature difference between the two bands, describes the real-space displacement encoded in the interband wave packet, and the average is weighted by the JDOS over the resonant states\cite{ma2021topology, ma2022intelligent, ahn2022riemannian, krishna2025terahertz, gao2025quantum}.

Shift currents have been observed in several quantum materials, including Weyl semimetals\cite{osterhoudt2019colossal, ma2019nonlinear, shao2021nonlinear, wang2023visualization}, BP/WSe$_2$ heterointerface\cite{akamatsu2021van}, 
and graphene moirés\cite{ma2022intelligent, krishna2025terahertz, persky2026revealing}.
However, many such quantum-geometric photovoltaics emerge at mid-infrared energy scales and, particularly in correlated systems, at terahertz frequencies.
At these frequencies, the optical wavelength exceeds the dimensions of typical devices, preventing spatial separation of bulk, contact, and interface contributions in far-field measurements. Higher-energy excitation improves spatial resolution but involves unwanted bands and relaxation pathways, obscuring the targeted low-energy response.

Here we introduce cryogenic, polarization-sensitive, terahertz microscopy for real-space imaging of quantum-geometric photovoltaics. Deep-subwavelength spatial resolution enables direct separation of bulk and contact contributions, while polarization control provides access to the real-space orientation and symmetry of the shift-current response. We apply this approach to dual-gated AB bilayer graphene, 
the thinnest member in the rhombohedral graphene family, in which electrostatic gating breaks inversion symmetry 
and continuously tunes electronic structure including its carrier density, band gap, and quantum geometry\cite{zhang2011spontaneous,sha2024observation,han2024large}.
When the band gap is tuned to approach the terahertz photon energy, we observe enhanced optical conductivity and BPVE, with a pronounced photocurrent extending throughout the device interior and an expected electron-hole asymmetry. Importantly, spatially resolved measurements further distinguish the intrinsic bulk response from interfacial photothermoelectric contributions. Furthermore, polarization- and temperature-dependent measurements reveal the evolution of the local shift-current response and its sensitivity to graphene crystallography and trigonal warping, in agreement with theoretical calculations. Our study not only uncovers the hidden Berry curvature and, in particular, quantum metric in AB bilayer graphene, but also reveals their tunability and sensitivity, offering an optical lens into the rich landscape of quantum geometry and electronic correlations in rhombohedral graphene and other quantum materials.

\vspace{-1ex}
\section*{Results}

\vspace{-2ex}
\subsection{Device and photoresponse}
\vspace{-1.0em}

Fig.~1a shows the device architecture and far-field THz measurement scheme. The active layer consists of AB-stacked bilayer graphene encapsulated by hBN, with graphite and monolayer graphene serving as the bottom and top gates, respectively. Independent control of the two gates enables continuous tuning of the carrier density $n$ and displacement field $D$, while the monolayer graphene top gate remains sufficiently transparent at THz frequencies to provide optical access to the bilayer graphene. The terahertz beam is focused onto the sample ($\lambda = 118~\mu$m, $\hbar\omega = 10.5$ meV).

We first consider the electronic structure of bilayer graphene and its expected shift-current response. Figure~1b shows the low-energy bands for an interlayer potential difference $\Delta=10.3$~meV, which is obtained by the full-band model with parameter values informed by our first-principles calculations. Because of the three-fold rotational ($C_{3z}$) and mirror ($M_x$) symmetries, the shift-current conductivity tensor has only one independent element, $\sigma_{yyy}$, proportional to the BPVE amplitude~\cite{ma2022intelligent}. Figure~1c shows the calculated $\sigma_{yyy}$ under $10.5$~meV excitation. The conductivity vanishes at $\Delta=0$ due to inversion symmetry and switches sign upon reversing $\Delta$ reflecting the odd parity of shift current under inversion. 

Moreover, The calculated response is strongly concentrated near the CNP, predominantly on the electron-doped side, and at $|\Delta|\simeq\hbar\omega$. The black contours in Fig.~1b denote the resonant states for the $\sigma_{yyy}$ maximum in Fig.~1c. Trigonal warping generates three off-center pockets related by $C_{3z}$, and the resonant contours pre-
\onecolumngrid
\vspace{-1ex}
\begin{center}
\includegraphics[width=0.92\textwidth]{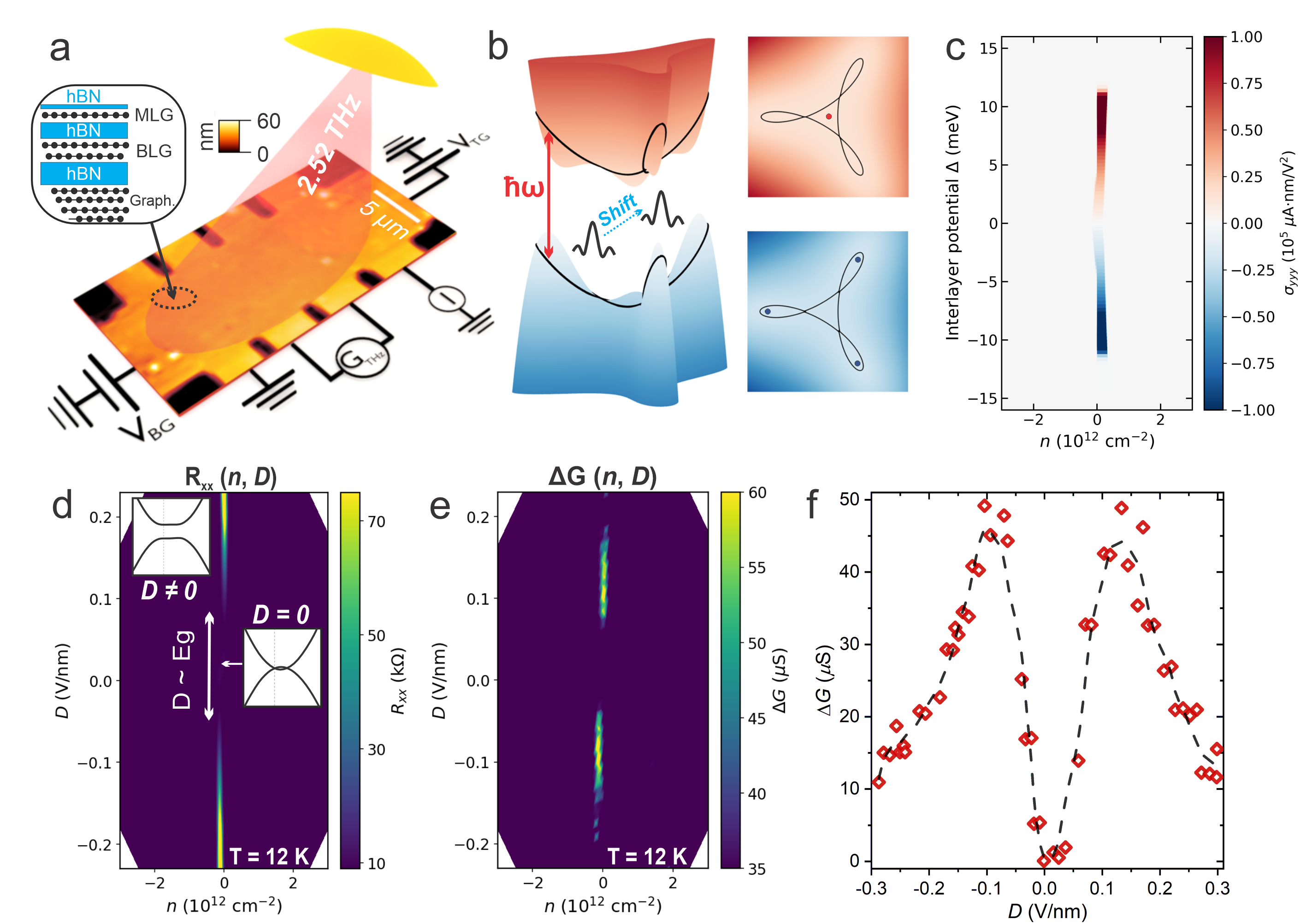}\\[0ex]
\parbox{1\textwidth}{\setlength{\parindent}{0pt}%
\textbf{Figure 1: Terahertz response of dual-gated AB-stacked bilayer graphene.}
\textbf{a}, Schematic of the device and far-field THz measurement. The bilayer graphene is encapsulated by hBN and controlled by top and bottom gates. The THz beam has a frequency of 2.52~THz ($\hbar\omega=10.5$~meV) and a spot size of approximately 0.5~mm. The inset shows the device cross section.
\textbf{b}, Calculated low-energy band structure of bilayer graphene at interlayer potential difference $\Delta=10.3$ meV. Black contours indicate resonant states satisfying $E_c(\mathbf{k})-E_v(\mathbf{k})=10.5$~meV. Red and blue dots mark the conduction-band minima and valence-band maxima, respectively.
\textbf{c}, Calculated shift-current conductivity $\sigma_{yyy}$ under 10.5~meV excitation. 
\textbf{d}, Longitudinal resistance $R_{xx}$ measured at 12~K. 
\textbf{e}, Change in conductance under THz illumination, $\Delta G=G_{\mathrm{laser\,on}}-G_{\mathrm{laser\,off}}$, measured at 12~K.
\textbf{f}, Linecut of $\Delta G$ along the CNP.}
\end{center}
\twocolumngrid
\noindent
dominantly sample these warped band-edge states. In this region, the JDOS is significantly enhanced by the presence of saddle points, and the quantum-geometric quantity $\left\langle g_{bb}R_a\right\rangle_{\omega}$ in Eq.~(\ref{sigma}) also peaks, thereby maximizing $\sigma_{yyy}$.
For weak electron doping, carriers initially populate the conduction-band minimum at the valley center, leaving the off-center resonant states largely unoccupied. Only at higher electron density, the Fermi level reaches the resonant states and Pauli blocking suppresses $\sigma_{yyy}$. On the hole-doped side, by contrast, the valence-band maxima are located at the three off-center pockets, and even weak hole doping rapidly depletes the states participating in the resonant transitions.

We next experimentally characterize the corresponding photoresponse. Figure~1d shows the longitudinal resistance $R_{xx}$ measured at 12 K. Increasing $|D|$ produces a pronounced enhancement of $R_{xx}$ at the CNP, consistent with the opening of a $D$-induced band gap. Figure~1e shows the change in conductance under THz illumination, $\Delta G=G_{\mathrm{laser\,on}}-G_{\mathrm{laser\,off}}$. Two pronounced regions of enhanced photoconductivity emerge near the CNP at $D\approx\pm0.1$~V/nm, where the band gap approaches $\hbar\omega$. Figure~1f shows a linecut of $\Delta G$ along the CNP. The two maxima occur when the $D$-induced band gap approaches the THz photon energy, providing evidence for an interband resonance. The $n$- and $D$-dependence reproduces the principal features of the calculated shift-current response in Fig.~1c, highlighting the sensitivity of the THz response to the trigonal-warped band-edge structure.

The far-field measurements establish both the relevant energy scale and the gate dependence of the optical response, while also highlighting the limitation of spatially integrating contributions from the entire device. Next, we perform local THz photocurrent measurements using cryogenic s-SNOM to spatially separate the bulk response from contact- and junction-related contributions and directly probe the quantum-geometric BPVE.

\vspace{-2ex}
\subsection{Spatial separation of BPVE and interfacial PTE}
\vspace{-1.0em}

Figure~2a illustrates the near-field scheme, in which the THz beam is focused onto a metallized AFM tip oscillating at $\Omega \approx 275$~kHz. The resulting local photocurrent is detected through the device contacts using lock-in demodulation at the tip oscillation frequency. Figure 2b shows the photovoltage amplitude measured more than $10~\mu$m away from the contacts as a function of $n$ and $D$. Two pronounced BPVE-generating regions emerge near the CNP on the lightly electron-doped side at $D\approx\pm0.1$~V/nm. Their positions nearly coincide with the resonant photoconductivity observed in Fig.~1c,e, including its pronounced electron--hole asymmetry. As discussed above, this asymmetry originates from the tri-
\onecolumngrid  
\vspace{-2.5ex}
\begin{center}
\includegraphics[width=0.93\textwidth]{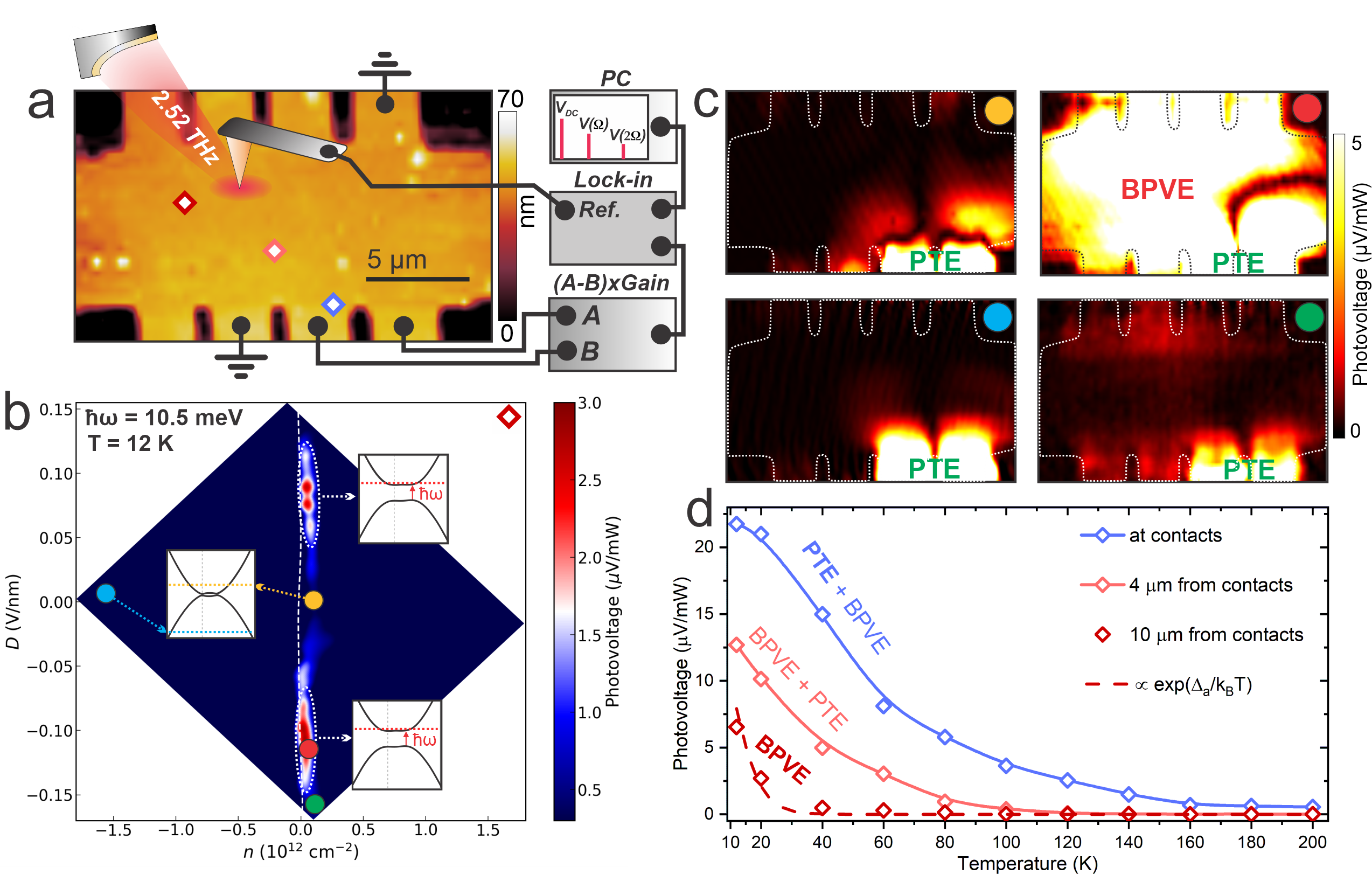} \\[0ex]
\parbox{1\textwidth}{\setlength{\parindent}{0pt}%
\textbf{Figure 2: Near-field measurements of quantum-geometric bulk photovoltaic effect at $12$~K under $10.5$~meV ($2.52$~THz) illumination.}  
\textbf{a}, AFM topography of the device and schematic of the near-field photocurrent measurement. 
\textbf{b}, Power-normalized local photovoltage amplitude as a function of displacement field $D$ and carrier density $n$, measured 10~$\mu$m away from the contacts (red diamond in panel \textbf{a}). The white dashed vertical line shows the position of CNP extracted directly from $R_{xx}$ measurements. Insets show the calculated band structures at the $(n, D)$ points indicated by the arrows.
\textbf{c}, Spatial maps of the local photovoltage measured at the four $(n, D)$ points indicated by the colored circles in panel \textbf{b}.
\textbf{d}, Temperature dependence of the power-normalized local photovoltage measured at the contacts (blue), 4~$\mu$m from the contacts (pink), and 10~$\mu$m from the contacts (red), as marked in panel \textbf{a}.}
\end{center}
\vspace{-1.0ex}
\twocolumngrid
\noindent
gonal-warped band-edge structure and provides further evidence that the observed response is governed by the shift-current mechanism. Upon increasing $|D|$ beyond the resonant condition, the band gap exceeds the THz photon energy and the bulk photovoltage rapidly vanishes, consistent with the interband origin.

The exceptional spatial selectivity of the measurement is further illustrated in Fig.~2c. Spatial maps acquired at representative points in the $(n,D)$ space reveal qualitatively different photocurrent distributions. Outside the BPVE-generating regions, including at $D=0$, the photoresponse is concentrated near contacts and gate-induced junctions, the characteristic of interfacial photo-thermoelectric effect (PTE). In contrast, within the BPVE-generating regions, a strong response extends throughout the device interior. This spatial selectivity allows the intrinsic bulk response to be identified directly rather than inferred from a signal containing inseparable contributions from multiple regions of the device.

The local nature of the measurement also enables temperature dependence to be resolved independently at different positions within the same device. Figure~2d shows the photovoltage measured at three representative locations: (i) at the contacts, where the response is dominated by the PTE (blue curve); (ii) approximately $4~\mu$m away from the contacts, where PTE and BPVE coexist (pink curve); and (iii) more than $10~\mu$m away from the contacts, where the response is governed by BPVE (red curve). In a far-field measurement, these spatially distinct contributions would be collected simultaneously and their different temperature dependences would be averaged together, obscuring the intrinsic bulk response.

The contact-dominated PTE signal (blue curve) increases upon cooling approximately as $T^{-1}$ before gradually saturating below $\sim30$~K, consistent with the reduced electron--phonon cooling and the resulting increase in the photoinduced electron temperature, $V_{\rm PTE}\propto S\Delta T_e$. By contrast, the BPVE signal in the bulk of the device (red curve) exhibits a markedly different behavior, decreasing exponentially with increasing temperature following $V_{\rm BPVE}\propto \exp\!\left({\Delta_a}/{k_BT}\right)$ with an activation energy of $\Delta_a\approx10$~meV obtained from the Arrhenius fit. This value closely matches the $D$-induced band gap, confirming that the local photovoltage originates from resonant interband transitions across the gap. The intermediate location (pink curve) exhibits a combination of both mechanisms, reflecting the gradual crossover from contact-dominated thermoelectric generation to the intrinsic bulk photovoltaic response.

\vspace{-2ex}
\subsection{Polarization dependence of local BPVE}
\vspace{-1.0em}

A smoking-gun approach for identifying shift-current-induced BPVE is to perform polarization-resolved measurements, as the response depends on the relative orientation between the incident linear polarization and the shift vector. 
As implied by Eq.~(\ref{sigma}), in a system with rotational symmetry, the photocurrent reaches its maximum 
when the polarization is aligned with the dominant shift-vector direction. 
In near-field experiments, however, the measured polarization dependence also contains an extrinsic contribution from the optical coupling geometry of the s-SNOM tip. A protocol for determining and subtracting this contribution is presented in Supplementary Information Section~III. In the following, we discuss the intrinsic polarization dependence of BPVE obtained from the local photovoltage signals after accounting for the extrinsic contribution.

Figure~3a illustrates the symmetry origin of the shift current and its polarization dependence. Owning to the three-fold rotational ($C_{3z}$) and mirror ($M_x$) symmetries, the shift vector $\bm R$ is constrained to lie along the mirror-line ($y$) direction. Consequently, the shift current along $y$ is maximized when the incident electric field $\bm E$ is either parallel or perpendicular to the mirror line, while it vanishes when the angle between them is $\pm 45^\circ$. The polarization dependence of the shift current along $x$ exhibits the opposite behavior. 

Figure~3b shows the polarization-dependent, power-normalized photovoltage measured in the bulk of the device for temperatures between 12 and 40~K. Upon cooling, the BPVE contribution rapidly increases and, below 40~K, closely follows the expected dependence~\cite{ma2022intelligent}, $V_{\rm BPVE}\propto\cos\!\left(2\theta-\phi\right)$, where $\theta$ denotes the linear polarization angle of the incident electric field and $\phi$ denotes the relative orientation between the voltage-collection direction and the mirror line of bilayer graphene. The excellent agreement between the experimental data and the theoretical expectation confirms that the observed photocurrent originates from the symmetry-constrained shift-current response. Additionally, this allows us to determine the crystallographic orientation of the device (Supplementary Information, Section III\textit{C}).

Figure~3c summarizes the temperature evolution of the polarization-dependent photovoltage. The BPVE remains robust between 12 and 40~K, while its amplitude rapidly decreases upon further warming, consistent with the temperature dependence observed in the spatially resolved measurements of Fig.~2d. This temperature sensitivity arises because the BPVE in bilayer graphene is governed by the resonant band gap and the trigonally warped band edges. Increasing the temperature can broaden the band-gap resonance and smear the band-edge occupations, thereby reducing the BPVE efficiency. 
\onecolumngrid
\vspace{-1ex}
\begin{center}
\includegraphics[width=1\textwidth]{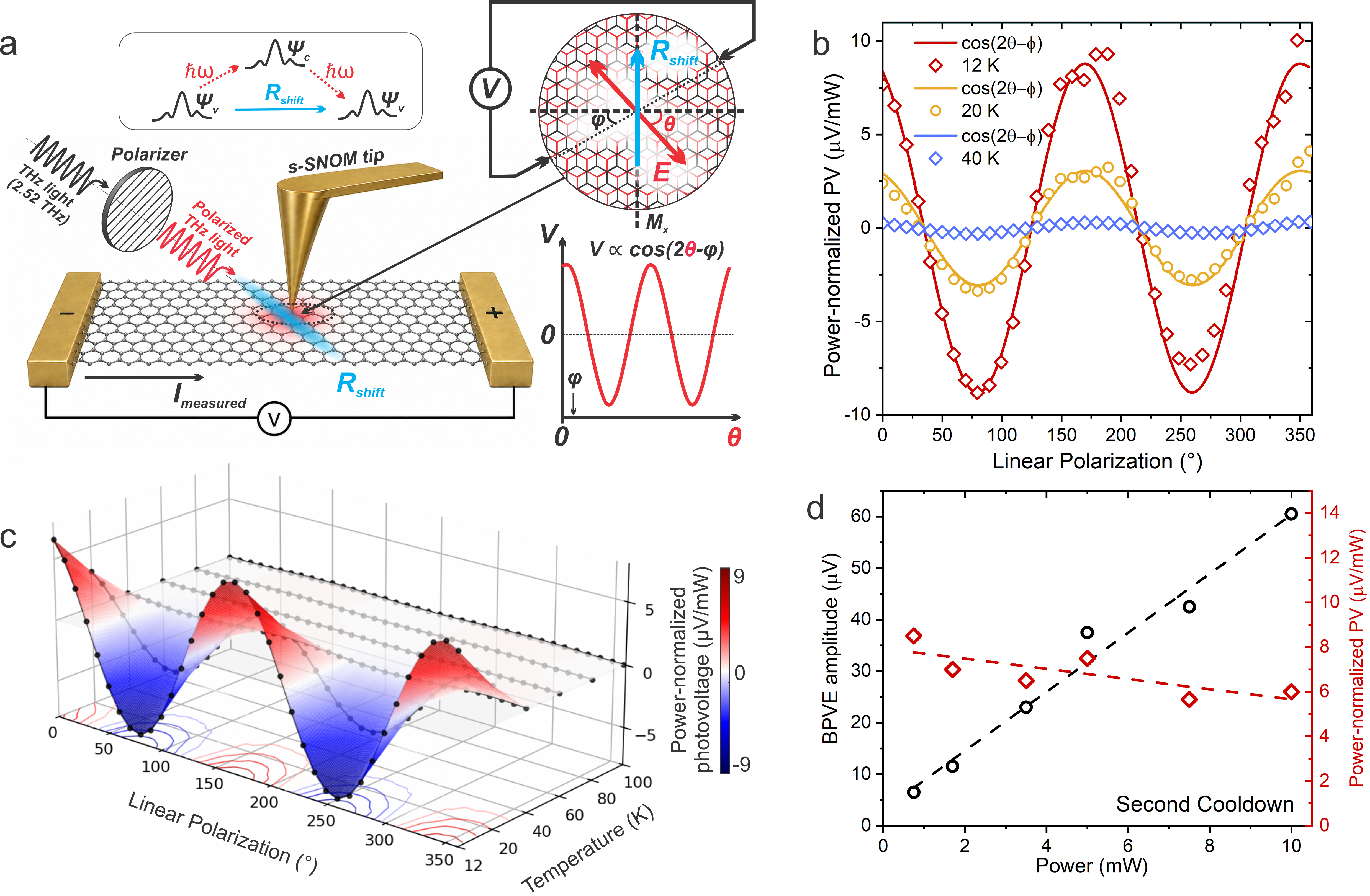}\\[0ex]
\parbox{1\textwidth}{\setlength{\parindent}{0pt}%
\textbf{Figure 3: Polarization dependence of local quantum-geometric bulk photovoltaic effect.}
\textbf{a}, Left: schematic of shift-current generation in bilayer graphene and measurement by s-SNOM. 
Upper right: $M_x$ mirror, shift vector $\bm R$ along $y$, and incident electric field $\bm E$. 
Lower right: Shift current $I_{\rm measured}$ as a function of the incident linear polarization angle $\theta$.  
\textbf{b}, Polarization-dependent, power-normalized photovoltage measured in the bulk of the device at temperatures between 40 and 12~K. The solid lines show the theoretically expected dependencies, $V_{\rm BPVE}\propto\cos(2\theta-\phi)$.
\textbf{c}, Temperature evolution of the polarization-dependent, power-normalized photovoltage.
\textbf{d}, Dependence of the photovoltage amplitude (black symbols, left axis) and the power-normalized amplitude (red symbols, right axis) on the
incident THz power. Dashed lines are guides to the eye.}
\end{center}
\twocolumngrid
Figure~3d shows the power dependence of the observed BPVE. The black symbols denote the photovoltage amplitude, which increases nearly linearly with excitation power, as expected for a second-order nonlinear effect. The red symbols denote the power-normalized photovoltage amplitude. Although this quantity would be expected to remain constant, we observe a weak decrease with increasing excitation power. We attribute this deviation to photo-induced electron heating. The strong near-field enhancement at the s-SNOM tip, which can amplify the local electric field by a factor of 10--100, can induce local electron heating even at low incident powers. As the incident power increases, so does the local electronic temperature. Thus, the power dependence of the power-normalized data is consistent with the temperature dependence observed in Fig.~3c. We note that the power-dependent measurements in Fig.~3d were performed after a second cooldown, and we observed a decrease in the BPVE amplitude following thermal cycling~\footnote{Such changes in the response are well known in van der Waals devices and can arise from modifications of the local mechanical configuration through thermal-expansion mismatch, strain redistribution, or relative relaxation between the constituent layers and contacts. Importantly, the qualitative response remained unchanged, supporting the robustness of the observed behavior.}.

\vspace{-2ex}
\section*{Discussion} 
\vspace{-1em}

Observation of BPVE at the nanoscale introduces challenges absent in conventional far-field measurements. In cryogenic s-SNOM, the metallic tip can itself act as a local heat source. We find that this contribution can be minimized by operating at high tapping setpoints (Supplementary Information, Section~II).
Despite the widespread use of s-SNOM for nanoscale optical measurements, polarization-resolved measurements of local photocurrents remain largely unexplored, because tip--sample coupling and tip geometry introduce their own polarization dependence. Here, we establish a protocol to separate the intrinsic polarization response from instrumental anisotropy (Supplementary Information, Section~III). 
Applying these approaches to dual-gated bilayer graphene, we observe the expected symmetry and phase of the shift current, 
validating polarization-resolved s-SNOM as a probe of local BPVE.

The resulting nanoscale sensitivity provides access to fine variations in the BPVE arising from the band geometry of bilayer graphene, including the band gap, carrier density, electron–hole asymmetry, trigonal warping, and crystallographic orientation. Energy scales of $\sim 1$~meV and carrier-density variations of $\sim 10^{11}$~cm$^{-2}$ can be resolved. 
Moreover, the measured BPVE and its polarization dependence provide direct signatures of the quantum geometry near the resonant band edges, revealing pronounced quantum metric contributions and identifying the dominant directions of the shift vector. 
Together, these measurements probe both the real and imaginary components of the quantum geometric tensor.

These capabilities are particularly relevant to quantum materials and devices characterized by small energy scales and strong spatial inhomogeneity. The combination of nanoscale spatial selectivity, cryogenic operation, electrostatic control, and polarization analysis established here provides a new route to probing intrinsic electronic properties that are difficult to disentangle in far-field measurements, including both band and quantum geometries.

\vspace{-1em}
\section*{\textsf{Methods}}
\small
\vspace{-1em}
\subsection*{Device fabrication.}
\vspace{-1em}
Si$^{++}$/SiO$_2$ substrates were pre-patterned using standard photolithography and metal deposition techniques. Photolithography was performed using a SUSS MicroTec MJB4 mask aligner and SPR-700 photoresist. After photoresist development, Ti/Au (10/50 nm) electrodes were deposited using an Angstrom Engineering electron-beam evaporation system.

hBN, graphite, and graphene flakes were prepared by mechanical exfoliation onto Si/SiO$_2$ substrates using adhesive tape. Suitable flakes were identified by optical microscopy based on their optical contrast. Van der Waals heterostructures were assembled using the well-established cut-and-stack technique\cite{saito2020independent}, which enables the fabrication of high-quality devices with atomically clean interfaces. The stacking sequence consisted of a thin hBN flake (3--7 nm), monolayer graphene serving as a THz-transparent top gate, an approximately 10--15 nm thick hBN spacer, AB-stacked bilayer graphene, a second 10--15 nm hBN spacer, and a graphite back gate. The completed heterostructure was transferred onto the pre-patterned Si$^{++}$/SiO$_2$ substrate using the PDMS/PC dry-transfer technique. 

Electron-beam lithography was performed using a Raith EBPG5200 system with PMMA 950 A4 resist. Following development, exposed regions were etched using an Oxford Instruments ICP-RIE system with CHF$_3$/O$_2$ plasma (40/6 sccm, 90 mTorr, 60 W). A second lithography step was used to define the contact geometry and connect the device to the pre-patterned electrodes. Finally, Cr/Au (2.5/30 nm) was deposited by electron-beam evaporation to form one-dimensional edge contacts.

\vspace{-1.5em}
\subsection*{Transport and optical conductivity measurements.}
\vspace{-1em}
\noindent
Electronic transport measurements were performed using a standard four-terminal low-frequency lock-in technique (17.777 Hz) with a Stanford Research Systems SR860 lock-in amplifier. The device was current biased with an AC current of $I_{\rm AC}=30$ nA. Top- and back-gate voltages were supplied using Keithley 2460 source meters.

Photoconductivity measurements were carried out by first recording resistance maps as a function of top- and back-gate voltages under dark conditions and subsequently under continuous-wave THz illumination provided by a commercially available FIRL-100 laser (Edinburgh Instruments). The photoconductivity was obtained by subtracting the dark conductivity map from the illuminated one. The THz beam was focused onto the sample using an off-axis parabolic mirror, producing a spot with a diameter of approximately 0.5 mm.

\vspace{-1.5em}
\subsection*{Near-field optical measurements.}
\vspace{-1em}
\noindent

Near-field measurements were performed using a commercially available cryogenic scattering-type scanning near-field optical microscope (cryo-neaSCOPE, Neaspec/Attocube). Local photovoltage was detected using lock-in demodulation at higher harmonics ($n\Omega$) of the AFM tip oscillation frequency, which suppresses the far-field THz background and isolates the deep-subwavelength near-field response. Similar approaches have previously been employed in Refs.~\cite{batlle2025photovoltage,hesp2024cryogenic,alonso2017acoustic}. Custom metal-coated AFM probes were illuminated by a continuous-wave FIRL-100 THz laser (Edinburgh Instruments). The photovoltage was measured using an Ithaco 1201 differential voltage preamplifier connected to the digital lock-in amplifier integrated into the cryo-neaSCOPE system. Polarization-dependent measurements were performed using a wire-grid polarizer (Pure Wave Polarizers) together with a THz half-wave plate (TYDEX).

\vspace{-1.5em}
\subsection*{THz tip fabrication.}
\vspace{-1em}
\noindent

Custom metal-coated AFM probes were fabricated for the THz near-field photovoltage measurements. Commercial silicon AFM probes (Nanoworld Arrow$^{\rm TM}$-NCR) with a nominal resonance frequency of approximately 250 kHz and a tip radius of approximately 10 nm were used as the starting templates. Prior to metallization, the probes were exposed to oxygen plasma (100 W, 30 s, 20 sccm O$_2$, 3 mTorr) to ensure the presence of a native oxide layer and improve metal adhesion. A 2 nm Ti adhesion layer was subsequently deposited by ion-assisted sputtering (50 W RF power, Ar flow 20 sccm), followed by the deposition of a 150 nm Au layer. All metallization steps were carried out using an Angstrom Engineering sputtering system.

\bibliography{References.bib}

\vspace*{-0.5cm}
\section*{\textsf{Acknowledgements}}
\vspace*{-0.2cm}
\noindent
We thank Justin C. W. Song for fruitful discussions. P.S. acknowledges support from the start-up fund provided by the University of Notre Dame (\#373837). P.S. acknowledge support from the National Science Foundation through grant No. 2542595. Conceptual work performed at UTD was support by the US Department of Energy under grant number DE-SC0026583 (F.Z.). Computational work performed at UTD was supported by the Welch Foundation under grant number AT-2264-20250403 (J.L.) and by the US National Science Foundation under grant numbers DMR-2414726 (F.Z.) and DMR-2324033 (F.Z.). J.L. and F.Z. acknowledge the Texas Advanced Computing Center (TACC) for providing supercomputing resources. K.W. and T.T. acknowledge support from the JSPS KAKENHI (Grant Numbers 21H05233 and 23H02052) and World Premier International Research Center Initiative (WPI), MEXT, Japan.

\vspace*{-0.5cm}
\section*{\textsf{Author contributions}}
\vspace*{-0.2cm}
{\small
\noindent
P.S., B.B., and F.Z. conceived and designed the experiment. B.B., X.Z., A.K., and S.G. fabricated samples and performed the measurements. The experimental data were analyzed and interpreted by B.B., R.K.K, and P.S. The theory and simulation work was done by J.L. and F.Z. The hBN crystals were provided by K.W. and T.T. B.B. and F.Z. wrote the manuscript, and all co-authors contributed to editing. P.S. supervised the work.
}

\vspace*{-0.2cm}
\section*{\textsf{Competing Financial Interests}}
\vspace*{-0.2cm}
{\small
\noindent
The authors declare no competing financial interests.
}

\vspace*{-0.2cm}
\section*{\textsf{Data Availability Statement}}
\vspace*{-0.2cm}
{\small
\noindent
The data that support the findings of this study is available from the corresponding authors upon reasonable request.
}
\newpage
\end{document}


\title{\Large\textsf{\papertitle}}

\author{Bogdan Borodin\equalcontrib}
\email{bborodin@nd.edu}
\affiliation{\footnotesize Department of Physics and Astronomy, University of Notre Dame, Notre Dame, IN 46556,~USA}
\affiliation{\footnotesize Stavropoulos Center for Complex Quantum Matter, University of Notre Dame, Notre Dame, IN 46556,~USA}

\author{Jiangxu Li\equalcontrib}
\affiliation{\footnotesize Department of Physics, University of Texas at Dallas, Richardson, TX 75080,~USA}

\author{~Xuanpu Zhang}
\affiliation{\footnotesize Department of Physics and Astronomy, University of Notre Dame, Notre Dame, IN 46556,~USA}
\affiliation{\footnotesize Stavropoulos Center for Complex Quantum Matter, University of Notre Dame, Notre Dame, IN 46556,~USA}

\author{Aleksei Karaborchev}
\affiliation{\footnotesize Department of Physics and Astronomy, University of Notre Dame, Notre Dame, IN 46556,~USA}
\affiliation{\footnotesize Stavropoulos Center for Complex Quantum Matter, University of Notre Dame, Notre Dame, IN 46556,~USA}

\author{Som Subhra Guria}
\affiliation{\footnotesize Department of Physics and Astronomy, University of Notre Dame, Notre Dame, IN 46556,~USA}
\affiliation{\footnotesize Stavropoulos Center for Complex Quantum Matter, University of Notre Dame, Notre Dame, IN 46556,~USA}

\author{Kenji Watanabe}
\affiliation{\footnotesize Research Center for Electronic and Optical Materials, National Institute for Materials Science, 1-1 Namiki, Tsukuba 305-0044,~Japan}

\author{Takashi Taniguchi}
\affiliation{\footnotesize Research Center for Materials Nanoarchitectonics, National Institute for Materials Science,  1-1 Namiki, Tsukuba 305-0044,~Japan}

\author{Roshan Krishna Kumar}
\affiliation{\footnotesize Catalan Institute of Nanoscience and Nanotechnology (ICN2), CSIC and BIST, Campus UAB, Bellaterra, Barcelona 08193, Spain}

\author{Fan Zhang}
\email{zhang@utdallas.edu}
\affiliation{\footnotesize Department of Physics, University of Texas at Dallas, Richardson, TX 75080,~USA}

\author{Petr Stepanov}
\email{pstepano@nd.edu}
\affiliation{\footnotesize Department of Physics and Astronomy, University of Notre Dame, Notre Dame, IN 46556,~USA}
\affiliation{\footnotesize Stavropoulos Center for Complex Quantum Matter, University of Notre Dame, Notre Dame, IN 46556,~USA}

\keywords{bulk photovoltaic effect, quantum geometry, bilayer graphene, shift current, s-SNOM}


\begin{abstract}
\vspace*{0.2cm}
\begin{quotation}
    \noindent\hspace*{-2.5em}%
    \small \textsuperscript{\#}These authors contributed equally to this work.
\end{quotation}
\end{abstract}

\maketitle
\clearpage
\onecolumngrid

\tableofcontents

\clearpage
\twocolumngrid


\renewcommand{\figurename}{Fig.}
\setcounter{equation}{0}
\setcounter{figure}{0}
\setcounter{table}{0}
\makeatletter
\renewcommand{\theequation}{S\arabic{equation}}
\renewcommand{\thefigure}{S\arabic{figure}}
\renewcommand{\thetable}{\arabic{table}}
\renewcommand{\bibnumfmt}[1]{[#1]}
\renewcommand{\citenumfont}[1]{#1}

\onecolumngrid

\section{Model Hamiltonian and shift-current response}
We describe the low-energy electronic structure of bilayer graphene using a
four-band continuum Hamiltonian near the valley $K$. In the sublattice basis
$(A_1,B_1,A_2,B_2)$, the Hamiltonian is
\begin{equation}
H(\mathbf{p}) =
\begin{pmatrix}
\Delta/2 & v_0 \pi^{\dagger} & v_4 \pi^{\dagger} & v_3 \pi \\
v_0 \pi & \Delta/2 & \gamma_1 & v_4 \pi^{\dagger} \\
v_4 \pi & \gamma_1 & -\Delta/2 & v_0 \pi^{\dagger} \\
v_3 \pi^{\dagger} & v_4 \pi & v_0 \pi & -\Delta/2
\end{pmatrix}.
\label{eq:blg_hamiltonian}
\end{equation}
Here $\Delta$ is the interlayer potential difference,
$\pi=p_x+i p_y$ is the complex momentum operator, with $p_x$ and $p_y$ the
$x$ and $y$ components of the momentum vector. The graphene lattice constant is
$a=0.246~\mathrm{nm}$, and
\begin{equation}
v_i=\frac{\sqrt{3}}{2}\frac{a\gamma_i}{\hbar},
\qquad i=0,3,4.
\end{equation}
The parameters $\gamma_0$, $\gamma_1$, $\gamma_3$, and $\gamma_4$ describe the
nearest-neighbor intralayer hopping, vertical interlayer hopping, and skew
interlayer hopping processes, respectively. At the $K$ point, the direct
low-energy gap is therefore set by the interlayer potential difference,
$E_g(K)=|\Delta|$.

\begin{figure}[h!]
    \centering
    \includegraphics[width=0.45\textwidth]{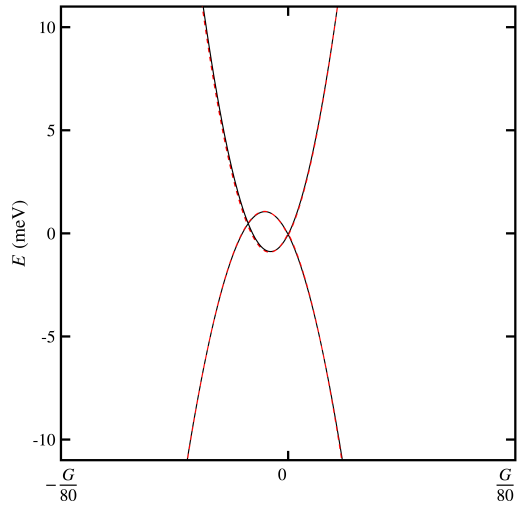}
    \caption{
    Comparison between the DFT band structure and the fitted four-band
    continuum model for bilayer graphene. The black solid curves denote the
    DFT bands, while the red curves denote the continuum-model bands obtained
    using the parameters in Eq.~\eqref{eq:blg_hamiltonian}. The excellent
    agreement near the Fermi level confirms that the chosen model parameters
    accurately reproduce the low-energy DFT electronic structure.
    }
    \label{fig:L2_DFTvTB}
\end{figure}

After validating the continuum model against the DFT band structure, we compute
the intrinsic shift-current response. For linearly polarized light with its
electric field polarized in the $\hat{j}$ direction, the intrinsic
shift-current conductivity tensor is expressed as
\begin{equation}
\sigma_{ijj}(0;\omega,-\omega)
=
\frac{\pi e^3}{\hbar^2}
\int_{\mathrm{BZ}} \frac{d^2 k}{(2\pi)^2}
\sum_{n,m}
f_{nm}\,
W^{ijj}_{nm}(\mathbf{k})
\delta(\omega_{nm}-\omega),
\label{eq:shift_current}
\end{equation}
where
\begin{equation}
W^{ijj}_{nm}(\mathbf{k})
=
\mathrm{Im}
\left[
r^j_{mn} r^i_{nm;j}
\right]
\label{eq:shift_quantum_geometric_weight}
\end{equation}
is the shift-current quantum-geometric weight. Here
$f_{nm}=f_n-f_m$ is the difference in Fermi occupation,
$r^i_{nm}=i\langle u_n|\partial_{k_i}|u_m\rangle$ is the non-Abelian Berry
connection, and
\begin{equation}
r^i_{nm;j}
=
\partial_{k_j} r^i_{nm}
-
i\left(r^j_{nn}-r^j_{mm}\right)r^i_{nm}
\end{equation}
is the generalized derivative. The transition energy is defined by
$\hbar\omega_{nm}=E_n-E_m$, and $\hbar\omega$ is the photon energy. The
integration is carried out over the first Brillouin zone.

At $\Delta=10.336~\mathrm{meV}$, the JDOS has a pronounced peak near
$10.5~\mathrm{meV}$, and the shift-current quantum-geometric weight is also
large in the same energy window, leading to an enhanced shift-current
conductivity.

\newpage
\section{Pure thermoelectric response in local s-SNOM measurements and its relation to the tapping conditions}

The photovoltage response measured in cryogenic s-SNOM experiments depends not only on the absorbed THz power but also on the mechanical interaction between the oscillating AFM tip and the sample surface. During the measurements, the AFM tip remains close to 250 K, whereas the graphene device is maintained at cryogenic temperatures (12 K). Consequently, whenever the oscillating tip comes into mechanical contact with the sample, heat is transferred from the warm tip into the graphene through the tip--sample junction. The amount of heat transferred during each oscillation cycle is therefore expected to depend primarily on the duration of mechanical contact.

To quantify this effect, we modeled the cantilever dynamics using a driven damped harmonic oscillator with nonlinear tip--sample interactions,

\vspace{-1.5em}
\begin{equation}
m\ddot{z}+\gamma\dot{z}+kz
=
F_{\rm drive}\cos(\omega_0t)
+
F_{\rm ts}(d),
\end{equation}
\vspace{-1.5em}

where $z(t)$ is the instantaneous tip displacement, $m$ is the effective cantilever mass, $\gamma$ is the damping coefficient, $k$ is the cantilever spring constant, $F_{\rm drive}$ is the excitation force, and $F_{\rm ts}$ describes the nonlinear tip--sample interaction.

The cantilever resonance frequency is

\vspace{-2em}
\begin{equation}
\omega_0=2\pi f_0,
\end{equation}
\vspace{-2em}

while the effective cantilever mass is

\vspace{-1.5em}
\begin{equation}
m=\frac{k}{\omega_0^2},
\end{equation}
\vspace{-1.5em}

and the damping coefficient is

\vspace{-1.5em}
\begin{equation}
\gamma=\frac{m\omega_0}{Q},
\end{equation}
\vspace{-1.5em}

where $Q$ is the cantilever quality factor.

The free oscillation amplitude was fixed at $A_0=250$~nm, while the tapping setpoint was defined as

\vspace{-1.5em}
\begin{equation}
SP=\frac{A_{\rm sp}}{A_0},
\end{equation}
\vspace{-1.5em}

where $A_{\rm sp}$ is the oscillation amplitude during scanning.

The instantaneous tip--sample separation is

\vspace{-2em}
\begin{equation}
d(t)=z(t)-z_{\rm surface}.
\end{equation}
\vspace{-2em}

For separations larger than the intermolecular distance ($d>a_0$), the interaction is described by the attractive van der Waals force,

\vspace{-1.5em}
\begin{equation}
F_{\rm vdW}
=
-\frac{HR}{6d^2},
\end{equation}
\vspace{-1.5em}

where $H$ is the Hamaker constant and $R$ is the tip radius.

Once the tip reaches the sample surface ($d\le a_0$), the interaction switches to the Derjaguin--Muller--Toporov (DMT) contact model,

\vspace{-1.5em}
\begin{equation}
F_{\rm rep}
=
\frac{4}{3}
E_{\rm eff}
\sqrt{R}
\,
\delta^{3/2},
\end{equation}
\vspace{-1.5em}

where $\delta=|d|$ is the indentation depth and $E_{\rm eff}$ is the effective elastic modulus. Adhesion is included through

\vspace{-1.5em}
\begin{equation}
F_{\rm adh}
=
-\frac{HR}{6a_0^2},
\end{equation}
\vspace{-1.5em}

such that

\vspace{-2em}
\begin{equation}
F_{\rm ts}=F_{\rm rep}+F_{\rm adh}.
\end{equation}
\vspace{-2em}

Mechanical contact is defined by

\vspace{-2em}
\begin{equation}
d(t)\le0.
\end{equation}
\vspace{-2em}

The total contact time accumulated during the simulation is

\vspace{-2em}
\begin{equation}
t_{\rm tot}
=
N_{\rm contact}\Delta t,
\end{equation}
\vspace{-2em}

where $N_{\rm contact}$ is the number of time steps satisfying the contact condition.

The fraction of the oscillation period spent in contact is

\vspace{-1.5em}
\begin{equation}
f_{\rm contact}
=
\frac{t_{\rm tot}}{t_{\rm sim}},
\end{equation}
\vspace{-1.5em}

and the average mechanical contact time per oscillation cycle is

\vspace{-2em}
\begin{equation}
t_{\rm contact}
=
f_{\rm contact}T,
\end{equation}
\vspace{-2em}

where

\vspace{-2em}
\begin{equation}
T=\frac{1}{f_0}
\end{equation}
\vspace{-2em}

is the cantilever oscillation period.

The nonlinear equation of motion was solved numerically using the implicit Radau integrator, which is well suited for the stiff dynamics introduced by the DMT interaction model.

\begin{figure}[h!]
    \centering
    \includegraphics[width=0.99\linewidth]{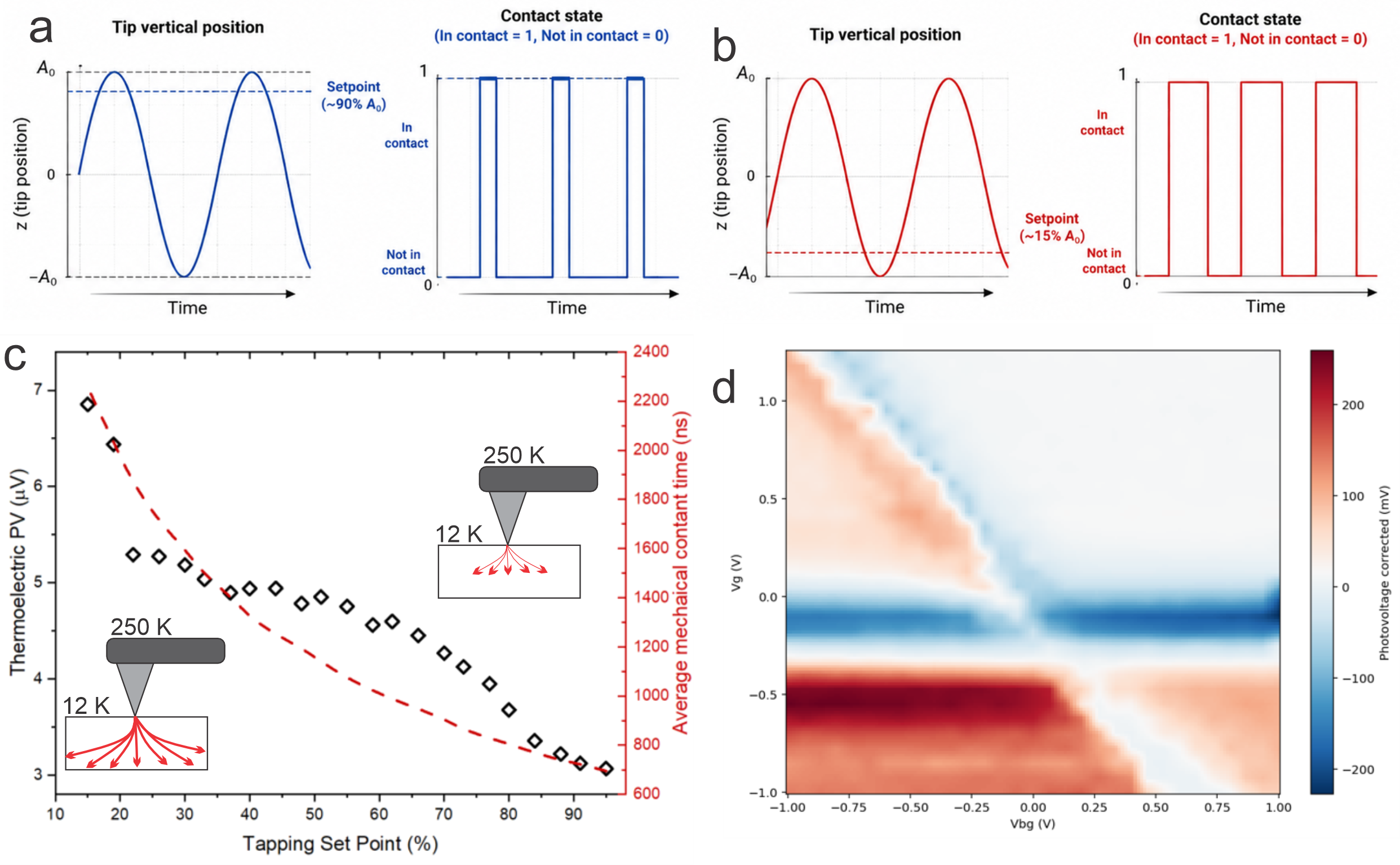}
    \vspace{-1em}
    \caption{\textbf{Dependence of the local thermoelectric response on AFM tapping conditions.}
    \textbf{a}) High tapping setpoint ($\sim90\%$ of the free oscillation amplitude). The tip contacts the sample only briefly during each oscillation cycle, resulting in a short mechanical contact time.
    \textbf{b}) Low tapping setpoint ($\sim15\%$). The tip remains in contact with the sample for a substantially larger fraction of the oscillation period.
    \textbf{c}) Local thermoelectric photovoltage measured at the device contacts as a function of tapping setpoint (black symbols). The red dashed curve shows the calculated average mechanical contact time obtained from the numerical cantilever simulation. The close correlation indicates that the thermoelectric response scales with the duration of tip--sample contact.
    \textbf{d}) Gate-dependent photovoltage measured at a low tapping setpoint ($\sim15\%$), where the response is dominated by the photothermoelectric effect. Strong signals appear at the charge-neutrality point and along electrostatic $pn$ junctions formed by the mismatch between the top- and bottom-gate potentials.}
    \label{fig:TE_response}
    \vspace{-1em}
\end{figure}

Figures~S2a and S2b illustrate the two limiting tapping regimes. At high tapping setpoints ($\gtrsim90\%$), the tip makes contact with the sample only during a very small fraction of each oscillation cycle, whereas lowering the setpoint to approximately $15\%$ substantially increases the mechanical contact time. The corresponding binary contact-state plots directly visualize this increase in the fraction of the oscillation period during which the tip physically touches the sample.

The experimentally measured thermoelectric response is presented in Fig.~S2c. The photovoltage increases monotonically as the tapping setpoint is reduced and closely follows the calculated average contact time obtained from the numerical cantilever dynamics. This strong correlation demonstrates that the local thermoelectric response is governed primarily by the duration of mechanical contact between the warm AFM tip and the cryogenic sample.

The physical origin of this behavior is the increased heat transfer through the tip--sample junction. Reducing the tapping setpoint increases the time during which thermal energy can be transferred from the hot tip into the graphene device, thereby increasing the local electron temperature and enhancing the thermoelectric voltage. Conversely, operating at high tapping setpoints minimizes the contact time and suppresses this parasitic heating mechanism.

Figure~S2d shows a representative gate-dependent photovoltage map acquired at a low tapping setpoint ($\sim15\%$), where the thermoelectric contribution dominates the measured signal. Pronounced photovoltage is observed at the charge-neutrality point and along electrostatic $pn$ junctions created by the mismatch between the top and bottom gates, consistent with the expected behavior of the thermoelectric effect. Throughout the main text, all BPVE measurements were therefore performed at tapping setpoints exceeding $90\%$ of the free oscillation amplitude in order to minimize tip-induced thermoelectric heating and isolate the intrinsic bulk photovoltaic response.

\newpage
\section{Extraction anisotropy in polarization-resolved s-SNOM photocurrent measurements}
\label{sec:extraction}
\subsection{Model of the coupling of polarized THz light to a s-SNOM tip}
In polarization-resolved s-SNOM photocurrent measurements, the measured angular dependence is not determined only by the intrinsic photocurrent generation mechanism. The incident THz field first couples to the metallic AFM tip, which acts as a local near-field antenna. Since the tip axis, illumination direction, and sample geometry define a preferred direction in the laboratory frame, the coupling efficiency depends on the angle between the incident linear polarization and this effective extraction axis.

For a linearly polarized incident electric field with polarization angle $\theta$, the electric-field projection onto the effective tip-coupling axis can be written as

\vspace{-1em}
\begin{equation}
E_{\rm loc}(\theta)
\propto
E_0 \cos(\theta-\phi_{\rm ext}),
\end{equation}

\noindent 
where $\phi_{\rm ext}$ is the angle of maximum coupling.
\begin{figure}[h!]
    \centering
    \includegraphics[width=0.4\linewidth]{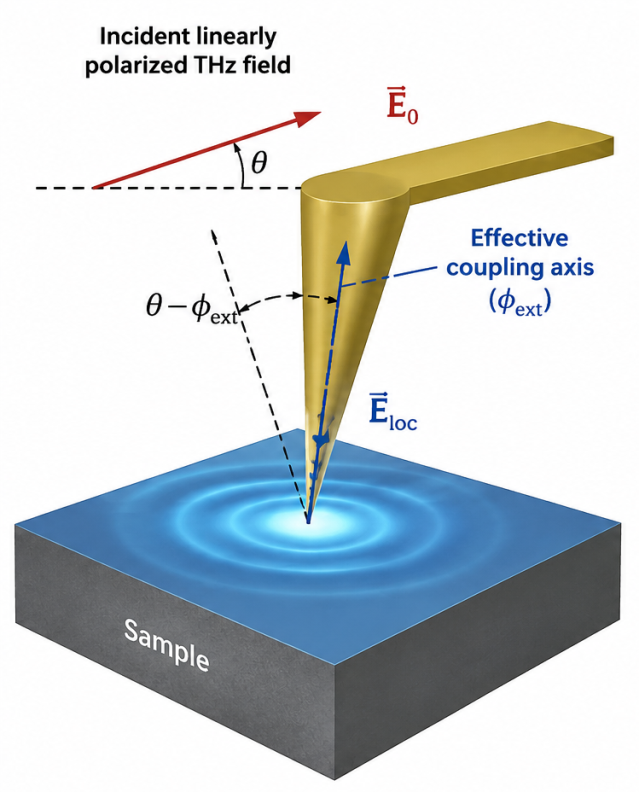}
   \caption{Schematic illustration of the coupling between the incident linearly polarized THz field and the metallic s-SNOM tip. The tip is approximated as a dipole antenna whose response is determined by the projection of the incident electric field onto the effective tip axis.}
    \label{fig:placeholder}
\end{figure}

Since the photocurrent is proportional to the absorbed optical power, and therefore to the square of the local electric field, the extraction efficiency scales as
\vspace{-0.5em}
\begin{equation}
\eta_{\rm ext}(\theta)
\propto
E_{\rm loc}^2(\theta)
\propto
\cos^2(\theta-\phi_{\rm ext}).
\end{equation}

Using
\vspace{-0.5em}
\begin{equation}
\cos^2 x
=
\frac{1+\cos(2x)}{2},
\end{equation}

\noindent   
the extraction anisotropy can be written, up to an overall amplitude factor, as
\vspace{-0.5em}
\begin{equation}
\eta_{\rm ext}(\theta)
=
1+\cos\!\left[2(\theta-\phi_{\rm ext})\right].
\label{eq:ext_function}
\end{equation}

Thus, even a photocurrent mechanism that is intrinsically independent of polarization will appear experimentally as a twofold function of the polarization angle. This is the central point of the correction procedure.

For example, the photothermoelectric contribution is not expected to depend intrinsically on the direction of linear polarization. However, because the measured signal is extracted through the polarization-dependent near-field coupling channel, the experimentally detected PTE signal is
\vspace{-0.5em}
\begin{equation}
I_{\rm PTE}^{\rm meas}(\theta)
=
A_{\rm PTE}
\left\{
1+\cos\!\left[2(\theta-\phi_{\rm ext})\right]
\right\}.
\label{eq:pte_measured}
\end{equation}

Therefore, the PTE-dominated region provides a direct experimental reference for determining the extraction phase $\phi_{\rm ext}$. In our analysis, the reference polarization trace is fitted using
\vspace{-0.5em}
\begin{equation}
I_{\rm ref}(\theta)
=
A_{\rm PTE}
\left\{
1+\cos\!\left[2(\theta-\phi_{\rm ext})\right]
\right\}.
\label{eq:reference_fit}
\end{equation}

This fit determines $\phi_{\rm ext}$, which is then kept fixed when analyzing the BPVE-active region.

The intrinsic bulk photovoltaic current generated by the shift-current mechanism has its own polarization dependence. For a dominant shift-vector direction $\phi_{\rm BPVE}$, the intrinsic BPVE response is modeled as
\vspace{-0.5em}
\begin{equation}
I_{\rm BPVE}^{\rm int}(\theta)
=
A_{\rm BPVE}
\left\{
C_{\rm BPVE}+\cos\!\left[2(\theta-\phi_{\rm BPVE})\right]
\right\}.
\label{eq:bpve_intrinsic}
\end{equation}

For bilayer graphene (BLG), which is a two-dimensional system with $M_x$ and $C_3$ symmetry, symmetry requires $C_{\rm BPVE}=0$. For other materials and crystal symmetries, however, $C_{\rm BPVE}$ may in general be nonzero.

However, this intrinsic BPVE current is detected through the same anisotropic s-SNOM extraction channel. Therefore, the measured BPVE contribution is not simply Eq.~(\ref{eq:bpve_intrinsic}), but rather the product of the intrinsic BPVE generation and the extraction efficiency:
\vspace{-0.5em}
\begin{equation}
I_{\rm BPVE}^{\rm meas}(\theta)
=
A_{\rm BPVE}
\left\{
C_{\rm BPVE}+\cos\!\left[2(\theta-\phi_{\rm BPVE})\right]
\right\}
\left\{
1+\cos\!\left[2(\theta-\phi_{\rm ext})\right]
\right\}.
\label{eq:bpve_measured}
\end{equation}

The total measured photocurrent in the BPVE-active region is then described as the sum of the PTE contribution, the detected BPVE contribution, and a polarization-independent offset:
\vspace{-0.5em}
\begin{align}
I_{\rm total}^{\rm meas}(\theta)
=
&\,
A_{\rm PTE}
\left\{
1+\cos\!\left[2(\theta-\phi_{\rm ext})\right]
\right\}
\nonumber\\
&+
A_{\rm BPVE}
\left\{
C_{\rm BPVE}+\cos\!\left[2(\theta-\phi_{\rm BPVE})\right]
\right\}
\left\{
1+\cos\!\left[2(\theta-\phi_{\rm ext})\right]
\right\}
+
C_0 .
\label{eq:total_model}
\end{align}

This is the fitting model used to decompose the measured polarization dependence. First, the PTE-dominated reference trace is fitted by Eq.~(\ref{eq:reference_fit}) to determine the extraction phase $\phi_{\rm ext}$. Then $\phi_{\rm ext}$ is fixed, and the total signal measured in the BPVE-active region is fitted using Eq.~(\ref{eq:total_model}), with $A_{\rm PTE}$, $A_{\rm BPVE}$, $\phi_{\rm BPVE}$, $C_{\rm BPVE}$, and $C_0$ as fitting parameters. For BLG, symmetry requires $C_{\rm BPVE}=0$, so this parameter is fixed to zero in the fit.

The product form in Eq.~(\ref{eq:bpve_measured}) is important because it naturally generates higher angular harmonics. In particular,
\vspace{-0.5em}
\begin{align}
&
\cos\!\left[2(\theta-\phi_{\rm BPVE})\right]
\cos\!\left[2(\theta-\phi_{\rm ext})\right]
\nonumber\\
&=
\frac{1}{2}
\cos\!\left[2(\phi_{\rm ext}-\phi_{\rm BPVE})\right]
+
\frac{1}{2}
\cos\!\left[4\theta-2(\phi_{\rm ext}+\phi_{\rm BPVE})\right].
\end{align}

Therefore, an intrinsically twofold BPVE response can appear in the raw measured data as a mixed twofold and fourfold polarization dependence. The fourfold component does not by itself imply a fourfold intrinsic BPVE symmetry; it can arise from the multiplication of the intrinsic twofold BPVE generation by the twofold extraction anisotropy. When $C_{\rm BPVE}\neq 0$, the additional term proportional to $C_{\rm BPVE}\cos[2(\theta-\phi_{\rm ext})]$ also contributes to the measured twofold component.

After fitting, the different components are reconstructed as
\vspace{-0.5em}
\begin{equation}
I_{\rm PTE}(\theta)
=
A_{\rm PTE}
\left\{
1+\cos\!\left[2(\theta-\phi_{\rm ext})\right]
\right\},
\end{equation}
\vspace{-0.5em}
\begin{equation}
I_{\rm BPVE}^{\rm meas}(\theta)
=
A_{\rm BPVE}
\left\{
C_{\rm BPVE}+\cos\!\left[2(\theta-\phi_{\rm BPVE})\right]
\right\}
\left\{
1+\cos\!\left[2(\theta-\phi_{\rm ext})\right]
\right\},
\end{equation}

and
\vspace{-0.5em}
\begin{equation}
I_{\rm BPVE}^{\rm int}(\theta)
=
A_{\rm BPVE}
\left\{
C_{\rm BPVE}+\cos\!\left[2(\theta-\phi_{\rm BPVE})\right]
\right\}.
\end{equation}

The last expression represents the intrinsic BPVE polarization dependence after removing the extraction anisotropy. In other words, the extracted BPVE response corresponds to the generation-level shift-current contribution before it is modulated by the polarization-dependent s-SNOM coupling efficiency.

In the implementation used here, the fitting was performed in two steps. First, the reference PTE trace was fitted to determine $\phi_{\rm ext}$. Second, the BPVE-active trace was fitted with the full model in Eq.~(\ref{eq:total_model}) using multiple random initial guesses to avoid convergence to a local minimum. The best solution was selected by minimizing the residual sum of squares between the measured and fitted total polarization traces. This procedure returns the PTE contribution, the BPVE contribution as detected through the anisotropic near-field channel, and the intrinsic BPVE contribution with the extraction anisotropy removed.

This correction is essential for interpreting polarization-resolved near-field photocurrent measurements. Without accounting for extraction anisotropy, a polarization-independent PTE response can be mistaken for a twofold intrinsic photocurrent, and an intrinsically twofold BPVE response can appear as a higher-harmonic angular pattern. By independently determining the extraction phase from a PTE-dominated reference region and then fixing it in the BPVE fit, the intrinsic shift-current polarization dependence can be separated from the experimental coupling geometry.

\subsection{Polarization-sensitive photovoltage data processing}
\vspace{-1em}
The extraction anisotropy, characterized by the phase $\phi_{\rm ext}$, can be determined experimentally using two independent approaches that produce consistent results.

The first approach exploits the thermal suppression of the BPVE. At sufficiently high temperatures, where the thermal energy satisfies $k_{\rm B}T \gtrsim E_g$, thermal broadening effectively washes out the displacement-field-induced band gap, strongly suppressing the quantum-geometric BPVE contribution. Under these conditions, the measured photovoltage is dominated by the photothermoelectric response, whose observed polarization dependence originates solely from the extraction anisotropy.

The second approach is performed at low temperature by tuning the carrier density $n$ and displacement field $D$ such that optical transitions at the excitation energy $\hbar\omega$ are blocked. In this regime, the BPVE is likewise suppressed, and the remaining polarization dependence again reflects only the polarization-dependent near-field coupling efficiency.

Figure~S4 compares polarization-dependent photovoltage traces obtained using these two approaches.

\begin{figure}[h!]
    \centering
    \vspace{-1em}
    \includegraphics[width=0.5\linewidth]{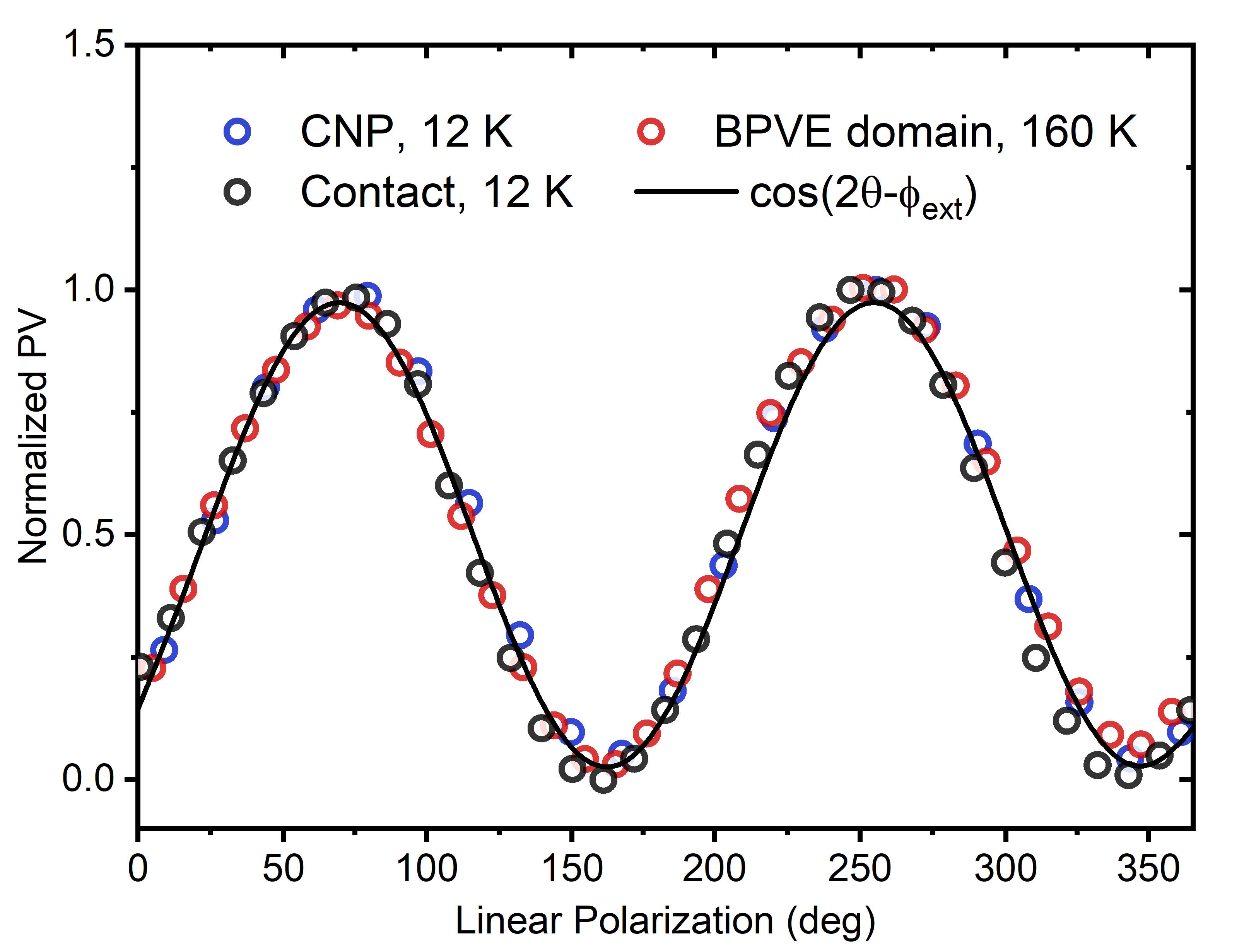}
       \vspace{-1em}
    \caption{\textbf{Extraction anisotropy determined using three independent experimental protocols.} (\textbf{i}) Polarization-dependent photovoltage measured at elevated temperature ($k_{\rm B}T \gtrsim E_g$), where the BPVE is thermally suppressed (red). (\textbf{ii}) Polarization-dependent photovoltage measured at low temperature under gate conditions that block resonant optical transitions at the excitation energy $\hbar\omega$ (blue). (\textbf{iii}) Polarization-dependent photovoltage measured at low temperature near the electrical contacts, where the response is dominated by the photothermoelectric effect (PTE) (green). Solid curves are fits to Eq.~(\ref{eq:reference_fit}), yielding the same extraction angle $\phi_{\rm ext}$ in all three cases.}
    \label{fig:SI5}
\end{figure}
\vspace{-1em}
In both cases, the experimental data are accurately described by the twofold angular dependence

\begin{equation}
I(\theta)
=
A\left[1+\cos\left(2(\theta-\phi_{\rm ext})\right)\right],
\end{equation}
yielding identical values of the extraction phase $\phi_{\rm ext}$ within the experimental uncertainty. This agreement confirms that the extraction anisotropy is governed by the optical measurement geometry and is independent of the electronic state of the bilayer graphene device. The extracted $\phi_{\rm ext}$ is subsequently used as a fixed reference in the decomposition of the polarization-dependent photovoltage into photothermoelectric and BPVE contributions.

Figure~S5 (top panel) illustrates the evolution of the polarization-dependent photovoltage as a function of temperature after entering the BPVE regime. As the temperature decreases below the energy scale set by the displacement-field-induced band gap, the polarization dependence progressively deviates from the simple twofold extraction anisotropy and develops the characteristic angular modulation associated with the intrinsic quantum-geometric BPVE. 

\begin{figure}[h!]
    \centering
    \vspace{-1.5em}
    \includegraphics[width=\linewidth]{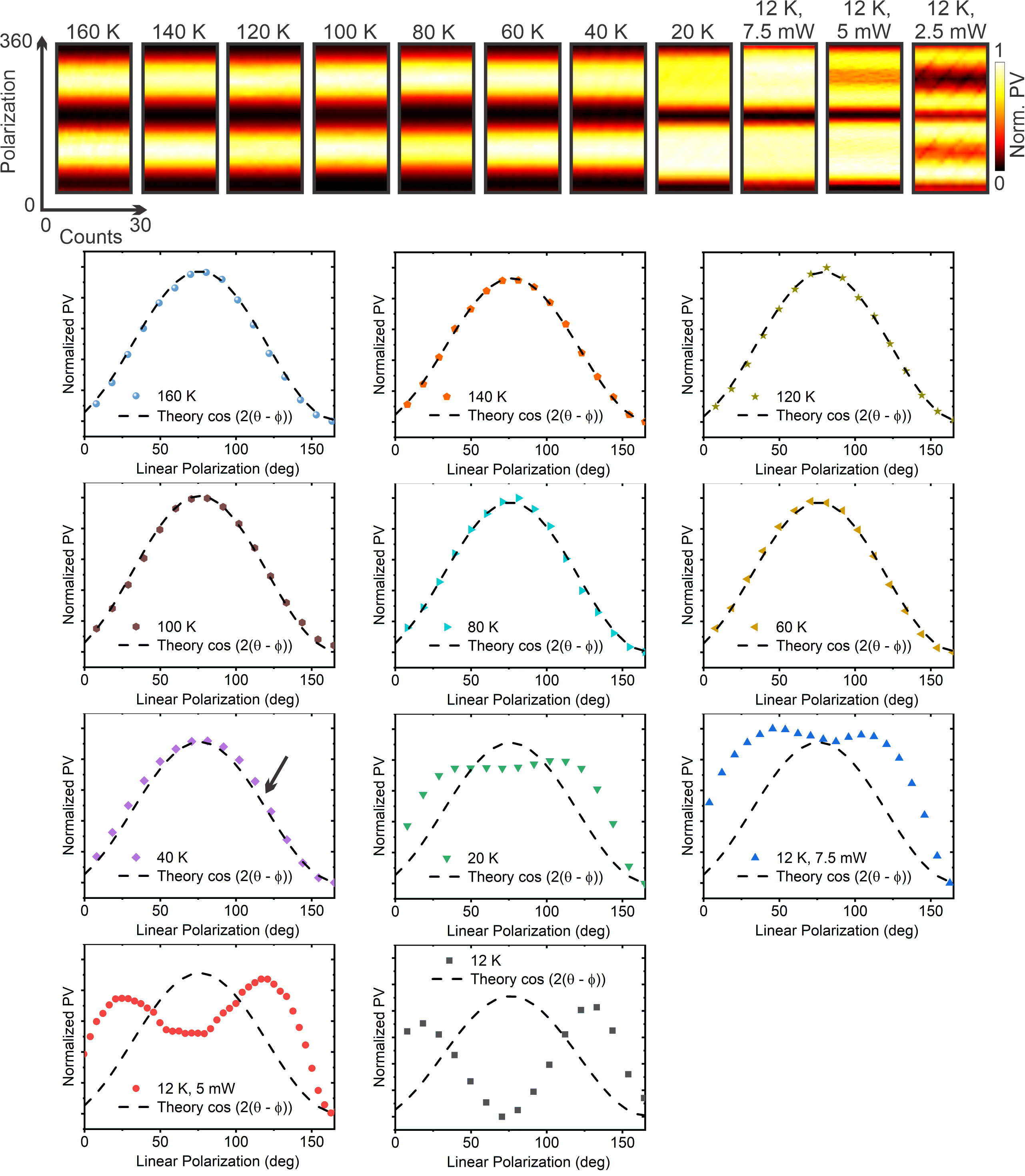}
    \vspace{-1.5em}
    \caption{\textbf{Temperature evolution of the polarization-dependent photovoltage within the BPVE domain.} \textbf{Top:} Raw polarization-resolved photovoltage maps plotted as a function of polarization angle and scan position for temperatures ranging from 160 to 12~K. At 12~K, measurements acquired at incident THz powers of 7.5, 5.0, and 2.5~mW are shown. \textbf{Bottom:} Corresponding normalized polarization traces together with fits to the twofold extraction function, $I(\theta)\propto1+\cos[2(\theta-\phi_{\rm ext})]$.}
    \vspace{-1.5em}
    \label{fig:SI3}
\end{figure}

Figure~S5 (bottom panel) summarizes the evolution of the polarization-dependent photovoltage with temperature. At temperatures down to approximately 60~K, the measured photovoltage is dominated by the photothermoelectric response and is accurately described by the twofold angular dependence expected from the extraction anisotropy, $I(\theta)\propto1+\cos[2(\theta-\phi_{\rm ext})]$.

At 40~K, small but reproducible deviations from the pure twofold behavior become apparent, as indicated by the arrow in Fig.~S5, signaling the onset of an additional polarization-dependent photocurrent contribution. Upon further cooling, these deviations become increasingly pronounced. At the lowest temperatures, where the displacement-field-induced band gap exceeds the thermal broadening, the polarization dependence clearly departs from the simple extraction function and develops the characteristic fourfold modulation discussed in Section~A. This evolution reflects the emergence of the intrinsic quantum-geometric BPVE, whose polarization dependence is superimposed on the fixed extraction anisotropy of the s-SNOM detection geometry.

The decomposition procedure described in the previous section is applied independently to each polarization trace in order to separate the photothermoelectric (PTE) and intrinsic BPVE contributions. For each dataset, the extraction phase $\phi_{\rm ext}$ is fixed to the value determined from the reference measurements described above, while the amplitudes of the PTE and BPVE contributions, the intrinsic BPVE phase $\phi_{\rm BPVE}$, and a constant offset are treated as free fitting parameters. To ensure convergence to the global optimum, each dataset is fitted using multiple random initial conditions, and the solution with the smallest residual sum of squares is selected.

\begin{figure}[h!]
    \centering
    \includegraphics[width=\linewidth]{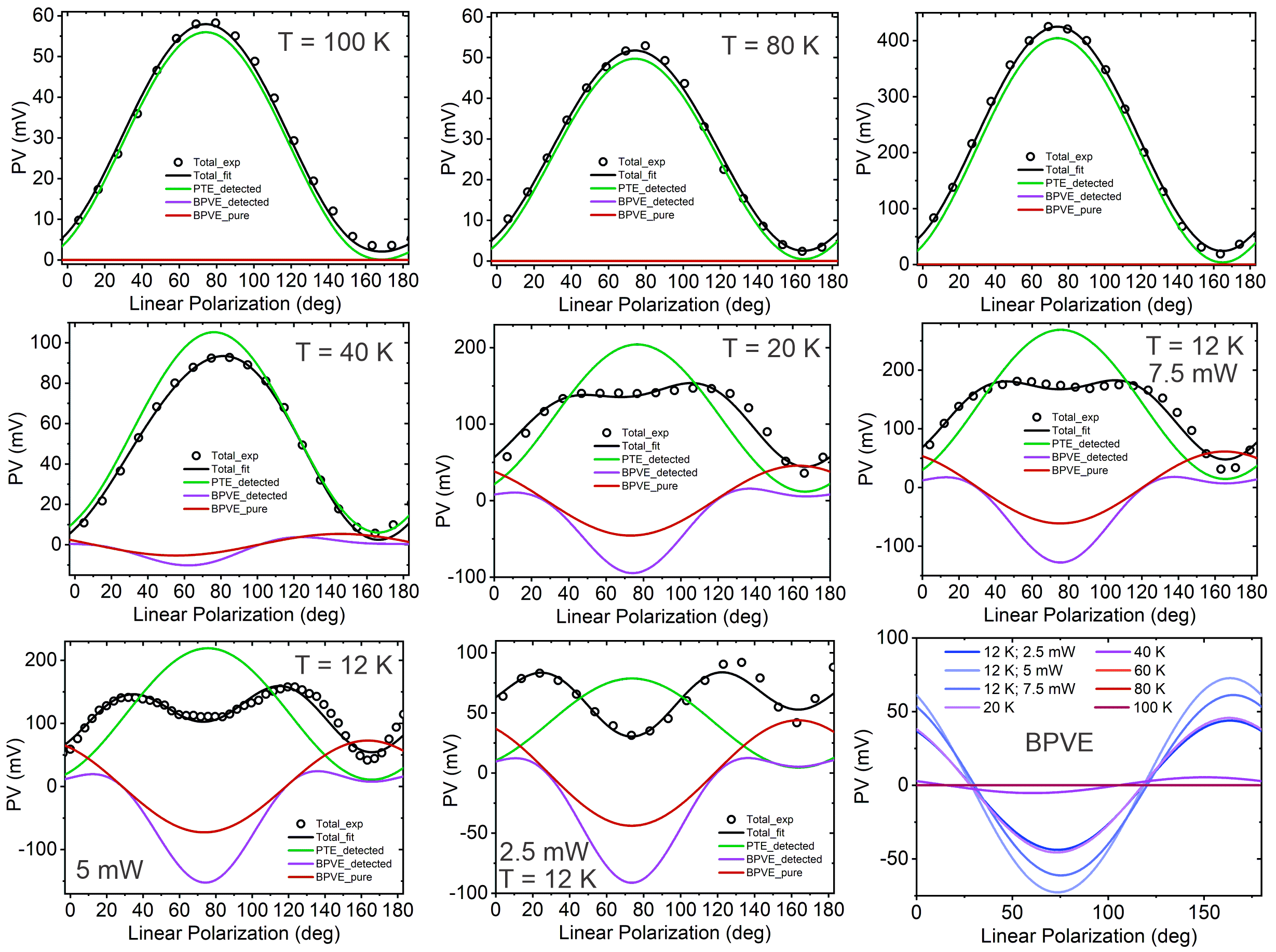}
    \caption{\textbf{Decomposition of polarization-dependent photovoltage into photothermoelectric and BPVE contributions.} Experimental polarization-dependent photovoltage (black open circles) together with fits to the complete model, Eq.~(S9) (black curves), for temperatures between 100 and 12~K. Measurements at 12~K were performed at incident THz powers of 7.5, 5.0, and 2.5~mW. For each dataset, the fitted photothermoelectric contribution (green), the detected BPVE contribution including the extraction anisotropy (purple), and the reconstructed intrinsic BPVE response after removing the extraction anisotropy (red) are shown separately. The bottom panel summarizes the intrinsic BPVE polarization dependences extracted from all datasets.}
    \label{fig:SI6}
\end{figure}

Figure~S6 demonstrates that the complete model provides an excellent description of the measured polarization dependence over the entire temperature range. At elevated temperatures, the experimental photovoltage is almost entirely described by the PTE contribution, while the fitted BPVE amplitude remains negligible. Consequently, the measured angular dependence closely follows the twofold extraction anisotropy determined from the reference measurements.

As the temperature is lowered below approximately 40~K, an additional polarization-dependent contribution becomes necessary to reproduce the experimental data. This contribution corresponds to the emergence of the quantum-geometric BPVE and progressively increases with decreasing temperature. At the lowest temperatures, the intrinsic BPVE becomes comparable to, and eventually exceeds, the PTE background, leading to the pronounced deviation from the simple twofold extraction function observed in the raw polarization traces.

The decomposition further reveals the influence of the excitation power on the relative strength of the two photocurrent mechanisms. At 12~K, reducing the incident THz power suppresses the photothermoelectric background considerably more rapidly than the BPVE contribution. As a result, the intrinsic BPVE becomes increasingly dominant, allowing its polarization dependence to be reconstructed with minimal influence from the extraction anisotropy.
    
The last panel of Fig.~S6 compares the intrinsic BPVE polarization dependences obtained from all measurements after removing the common extraction anisotropy. Despite the substantial variation in temperature and excitation power, all reconstructed traces exhibit the expected twofold symmetry characteristic of shift-current generation. The progressive increase in their amplitude with decreasing temperature and optical power is consistent with the enhancement of the quantum-geometric BPVE discussed in the main text.

\subsection{Real-Space Dominant Shift-Vector Orientation}

\begin{figure}[h!]
\centering
\includegraphics[width=\linewidth]{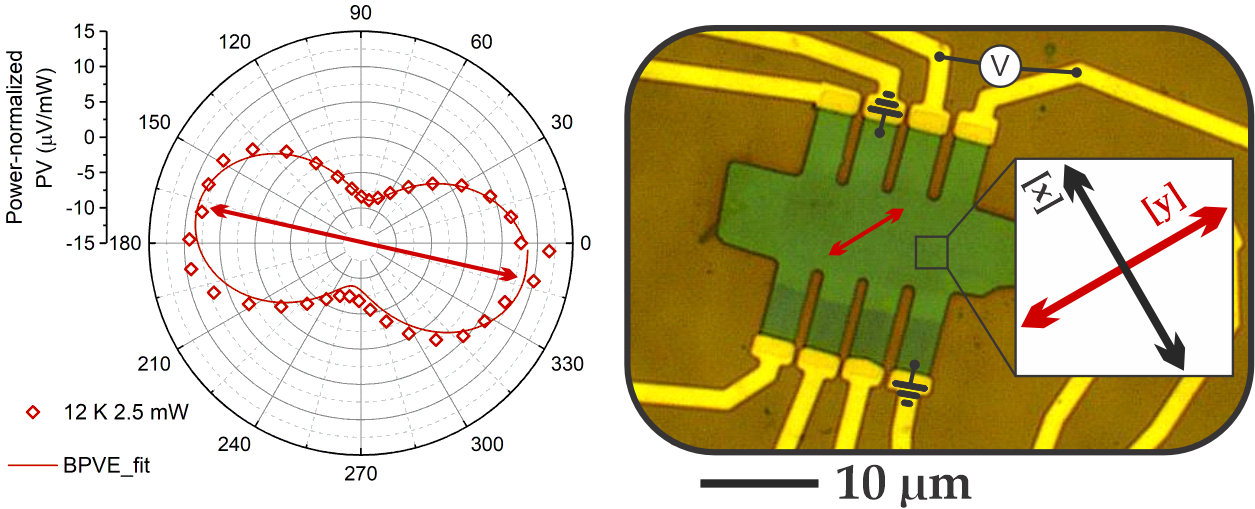}
\caption{\textbf{Real-space dominant shift-vector orientation and associated crystallographic directions.} The left panel shows the relative orientation of the incident polarization plane, with $0^\circ$ corresponding to $p$-polarization. The right panel shows an optical image of the device, with the direction of maximum photovoltage indicated by the red arrow. The corresponding $[y]$ and $[x]$ crystallographic directions are assigned based on the polarization dependence of the photovoltage.}
\label{fig:SI8}
\end{figure}

As indicated by the symmetry analysis, the maximum shift-current contribution is expected along the $[y]$ ($M_x$ mirror axis) direction. Accordingly, the orientation of the photovoltage maximum obtained from polarization-resolved measurements provides a direct indication of this crystallographic direction. The left panel of Fig.~\ref{fig:SI8} shows the relative orientation of the incident polarization plane, with $0^\circ$ corresponding to $p$-polarization. The right panel shows an optical image of the device, with the direction of maximum photovoltage indicated by the red arrow. Based on the observed polarization maximum, we assign this direction to the $[y]$ crystallographic axis and define the orthogonal direction as $[x]$, as indicated in the figure.

